\documentclass[prl,showpacs,english,superscriptaddress,twocolumn]{revtex4}
\usepackage{amsmath,amssymb,amsfonts,mathrsfs}
\usepackage{amsthm}
\usepackage{CJK}
\usepackage{bm}
\usepackage{babel}
\usepackage{graphicx}

\begin{document}

\title{Ferromagnetic Dirac-Nodal Semimetal Phase of Stretched Chromium Dioxide}

\author{R. Wang$^{*}$}
\affiliation{Department of Physics, South University of Science and Technology of China,
Shenzhen 518055, P. R. China.}
\affiliation{Institute for Structure and Function $\&$
Department of physics, Chongqing University, Chongqing 400044, P. R. China.}

\author{J. Z. Zhao$^{*}$}
\affiliation{Department of Physics, South University of Science and Technology of China, Shenzhen 518055, P. R. China.}
\affiliation{Dalian Institute of Chemical Physics,Chinese Academy of Sciences, 116023 Dalian, P. R. China}
\author{Y. J. Jin}
\affiliation{Department of Physics, South University of Science and Technology of China, Shenzhen 518055, P. R. China.}
\author{Y. P. Du}
\affiliation{Department of Applied Physics, Nanjing University of Science and Technology, Nanjing,
 210094 Jiangsu, P. R. China}
\author{Y. X. Zhao$^\dag$}
\affiliation{Department of Physics and Center of Theoretical and Computational Physics, The University of Hong Kong,
	Pokfulam Road, Hong Kong, China.}
\author{H. Xu$^\dag$}
\affiliation{Department of Physics, South University of Science and Technology of China, Shenzhen 518055, P. R. China.}
\author{S. Y. Tong$^\dag$}
\affiliation{Department of Physics, South University of Science and Technology of China, Shenzhen 518055, P. R. China.}
\affiliation{School of Science and Engineering, The Chinese University of Hong Kong (Shenzhen), 518172 Shenzhen, P. R. China}

\begin{abstract}
We show by first-principles calculations that the Dirac nodal-line semimetal phase can co-exist with the ferromagnetic order at room temperature in chromium dioxide, a widely used material
in magnetic tape applications, under small tensile hydrostatic strains. An ideally flat Dirac nodal ring close to the Fermi energy is placed in the reflection-invariant boundary of the Brillouin zone
perpendicular to the magnetic order, and is topologically protected by the unitary mirror symmetry of the magnetic group $D_{4h}(C_{4h})$, which quantizes the corresponding Berry phase into integer multiples
of $\pi$. The symmetry-dependent topological stability is demonstrated through showing that only the topologically protected nodal ring can persistently exist under small anisotropic stains preserving
the symmetry $D_{4h}(C_{4h})$, while the other seeming band touching points are generically gapped.
Our work provides a practical platform for the investigation of novel physics and potential applications of the Dirac nodal-line and drumhead fermions, in particular those related to ferromagnetic properties.
\end{abstract}

\maketitle

Over the past decade, the discovery of topological insulating and semi-metallic materials with electronic band structures protected by the interplay of symmetry and topology has drawn broad interest in condensed matter physics \cite{Hasan2010, Qi2011, Hsieh2012, Burkov2011, Young2012, Fang2015, Jeon2014,Liu2014}.
Compared to ordinary three-dimensional (3D) metals in which the filled and empty states are separated by two-dimensional (2D) Fermi sheets, 3D topological semimetals exhibit zero-dimensional (0D) discrete nodal points or one-dimensional (1D) continuous nodal lines \cite{Burkov2011}. In Weyl semimetals \cite{Xu2015,Lv2015,Yang2015}, the nodal points host nonzero chiral charges that are connected by Fermi arc surface states \cite{Wan2011,Xu2011}. Dirac node-line semimetals (DNLMSs), on the other hand, possess 1D Fermi surfaces accompanied by drumhead surface states \cite{Burkov2011}. Recently topological gapless modes of DNLs have been demonstrated to support unusual transport properties~\cite{Rui-Zhao-Schnyder-DNL-Trans}, and much attention has been focused on the exotic correlation physics of DNLs \cite{PhysRevB.83.220503,NaturePhysics10.964}. Unlike Weyl semimetals, the stability of a DNLSM requires symmetry, which may be either a space-time inversion $IT$ with spin-rotation symmetry~\cite{Zhao-Schnyder-Wang-PT-Classification,Fang2015} or a mirror reflection $M$~\cite{Chiu-Reflection-Classification,Fang2015,Zak-phase}. For the $IT$-protected DNLSMs, the spin-rotation symmetry enables us to effectively treat electrons as spinless fermions with $(IT)^2=1$, corresponding to the $\mathbb{Z}_2$ classification~\cite{Zhao-Schnyder-Wang-PT-Classification}. DNLs of this kind can be gapped by SOCs, since SOCs break the spin-rotation symmetry, and therefore lead to $(IT)^2=-1$ making the symmetry protection of the topological charge ineffective. Due to the spin degeneracy, each point in such a nodal line has the four-fold degeneracy. Examples are band crossings in Cu$_3$NZn and Cu$_3$NPd \cite{Kim2015, Yu2015prl}, Ca$_3$P$_2$~\cite{Xie2015, Chan2016}, and hyper honeycomb structures~\cite{Mullen2015}. On the other hand, a DNL in a mirror plane in the first Brillouin zone (BZ) can also be protected by the unitary mirror symmetry $M$~\cite{Chiu-Reflection-Classification,Fang2015}, which has two mutually opposite eigenvalues. When valence bands inside and outside a DNL have opposite eigenvalues of $M$, the DNL is topologically protected by the mirror symmetry with the two-fold degeneracy. From the topological viewpoint, a unitary mirror symmetry can quantize the Berry phase for a continuous band along a mirror-symmetric circle~\cite{Zak-phase}. In principle, mirror-symmetry-protected DNLSMs do not require the inversion or time-reversal symmetry, and may have strong spin-orbital couplings. While noncentrosymmetric materials with strong spin-orbital couplings, for instance PbTaSe$_2$, and TlTaSe$_2$ \cite{ Ali2014, Bian2016nc, Bian2016prb}, have been studied, DNLSMs with magnetic order breaking the time-reversal symmetry are still awaited to be explored. In addition, most candidates of DNLSMs thus far requires low temperature and do not have ideally flat DNLs, which serve as disadvantages for experimentally testing the novel physics and potential applications.

In this Letter, based on first-principles calculations with the symmetry and topology analysis, we propose that the cubic chromium dioxide (CrO$_2$) under small tensile hydrostatic strains can host a ferromagnetic (FM) DNLSM phase at room temperature, where a single topological Dirac nodal ring (DNR) on the reflection-invariant boundary of the Brillouin zone (BZ) perpendicular to the FM order is protected by the corresponding mirror symmetry. As a significant advantage, this DNR appears to be nearly flat and very close to the Fermi level, providing a more ideal platform for exploring novel DNL physics compared with other existing candidates~\cite{Yu2015prl,Bian2016nc,Bian2016prb}. It is noteworthy that CrO$_2$ is a very common material in practice. For instance, the rutile phase of CrO$_2$ with a high Curie temperature of about 390K is widely used in data tape applications~\cite{Schwarz1986,Katsnelson2008}. It has been demonstrated that, with a critical external pressure of $\sim$88.8 GPa, the cubic phase is able to be obtained from the rutile one and keep stable in a very large pressure range~\cite{Kim2012}.
Above the Curie temperature the cubic CrO$_2$ crystal has the $O_h$ symmetry as the point group of the symmorphic crystallographic group $Fm\bar{3}m$ (No. 225) and the time-reversal symmetry $T$, which are reduced to the magnetic group $D_{4h}(C_{4h})$ after lowering the temperature below the Curie temperature with the FM order being developed along any one of the three equal principle axes. After exerting tensile strains preserving the magnetic group $D_{4h}(C_{4h})$, it is found that the tetragonal CrO$_2$ in the presence of SOC  exhibits in its band structure a single nodal ring on the boundary of the BZ perpendicular to the FM order. It is observed that the magnetic group $D_{4h}(C_{4h})$ has only one unitary mirror symmetry with the mirror plane perpendicular to the FM order, and the DNL lies in the BZ boundary invariant under the reflection. Remarkably, due to the magnetic order breaking both $T$ and spin rotation symmetries, the nodal line has the two-fold degeneracy protected by the unitary mirror symmetry, which remains stable even in the presence of strong SOC. The Berry phase of valence bands along any mirror-symmetric circle surrounding the DNR is quantized by the mirror symmetry~\cite{Zak-phase}. Consequently, the Berry phases of 1D gapped subsystems  (parallel to the magnetic order) inside the nodal ring are nontrivial, while those outside are trivial, leading to the drumhead states covering the inner region of the projection of the DNR into the surface BZ parallel to the DNR or perpendicular to the FM order.



\begin{figure}
	\centering
	\includegraphics[scale=0.056]{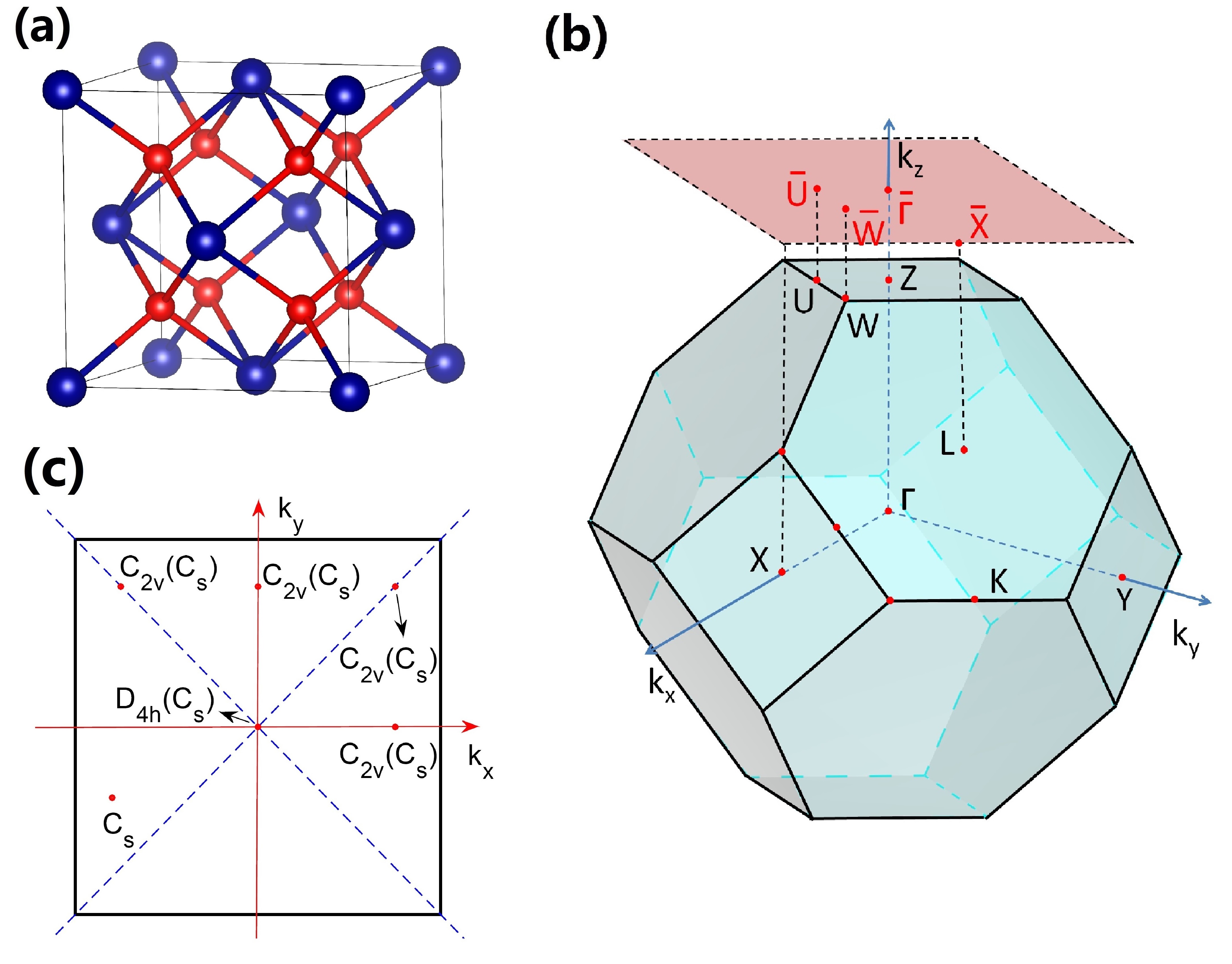}
	\caption{(a) Crystal structure of cubic CrO$_2$ with the space group $Fm\bar{3}m$ (No. 225). Cr and O atoms are indicated by blue and red spheres, respectively. (b) The FCC BZ and the corresponding (001) surface BZ. The independent $T$-invariant points are $\Gamma$(0, 0, 0), $L (\pi/a, \pi/a, \pi/a)$, $X (2\pi/a, 0, 0)$, $Y (0, 2\pi/a, 0)$, and $Z (0, 0, 2\pi/a)$. (c) Little groups for points in the BZ boundary of $k_z=2\pi/a$ \label{figure1}}.
\end{figure}


The first-principles calculations are carried out using density functional theory (DFT)~\cite{Hohenberg, Kohn} as implemented in the Vienna Ab initio Simulation Package (VASP)~\cite{Kresse2, Kressecom}. The core-valence interactions are treated by the projector augmented wave (PAW) method~\cite{Blochl,Kresse4, Ceperley1980}, where the plane wave expansion is truncated with a cutoff energy of 600 eV. We employ the exchange-correlation functional as the generalized gradient approximation (GGA) with the Perdew-Burke-Ernzerhof (PBE) formalism \cite{Perdew1, Perdew2}. In self-consistent calculations, a dense k-mesh with a  $35 \times 35 \times 35$ Monkhorst-Pack grid in momentum space \cite{Monkhorst} has been used to determine the gap of SOC. We introduce the on-site Coulomb repulsion beyond the GGA, i.e., GGA+U calculations \cite{Liechtenstein1995, Korotin1998} to fit the strongly correlated effects of 3d electrons in chromium. The value of correlation energy is chosen to be 3.0 eV, which works well in fitting the half-metallic properties of rutile CrO$_2$ \cite{Yamasaki2006, PhysRevLett.80.4305}.

The cubic CrO$_2$ crystalizes in a face-centered-cubic (FCC) lattice as shown in Fig. \ref{figure1}(a), which has the symmorphic space group $Fm\bar{3}m$ (No. 225) that is the semidirect product of the point group $O_{h}$ with the group $\mathbb{Z}^3$ generated by three lattice translations. The crystal structure consists of interpenetrating Cr and O sublattices, where the Cr atom is positioned at $(0, 0, 0)$, and O atoms are placed at $(0.25, 0.25, 0.25)$ and $(0.75, 0.75, 0.75)$, respectively. The FCC BZ and corresponding $(001)$ surface BZ are shown in Fig. \ref{figure1}(b) with high-symmetry points being indicated.

\begin{figure}
	\centering
	\includegraphics[scale=0.09]{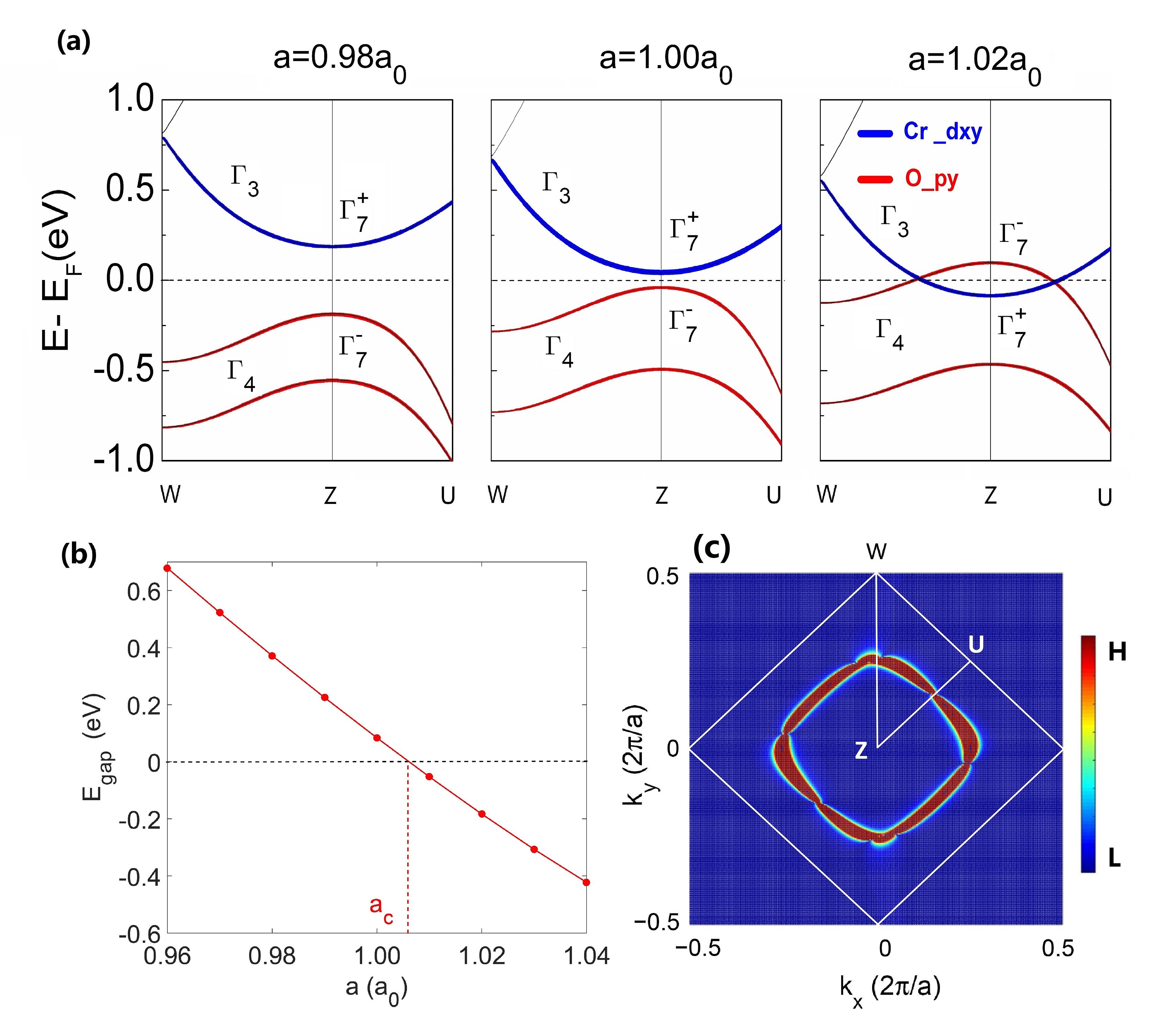}
	\caption{(a) Formation of the DNR from band inversion induced from increasing the tensile hydrostatic strain in GGA+U+SOC calculation. The orbital resolved band structures is along W-Z-U, and relevant irreducible representations are indicated. The component of Cr$\_d_{xy}$ (O$\_p_{y}$) orbitals is proportional to the width of the blue (red) curves.
	(b) Energy gap of cubic CrO$_{2}$ at the Z point as a function of lattice constant (red) with U=3.0 eV. A negative gap implies that the $\Gamma_{7}^{+}$ and $\Gamma_{7}^{-}$ states with opposite mirror eigenvalues are inverted and a nodal ring has been formed. (c) The Berry curvature at the $k_{z}=2\pi/a$ plane diverges along the nodal line and vanishes away from the DNR. \label{figure2}}
\end{figure}

\begin{figure}
	\centering
	\includegraphics[scale=0.090]{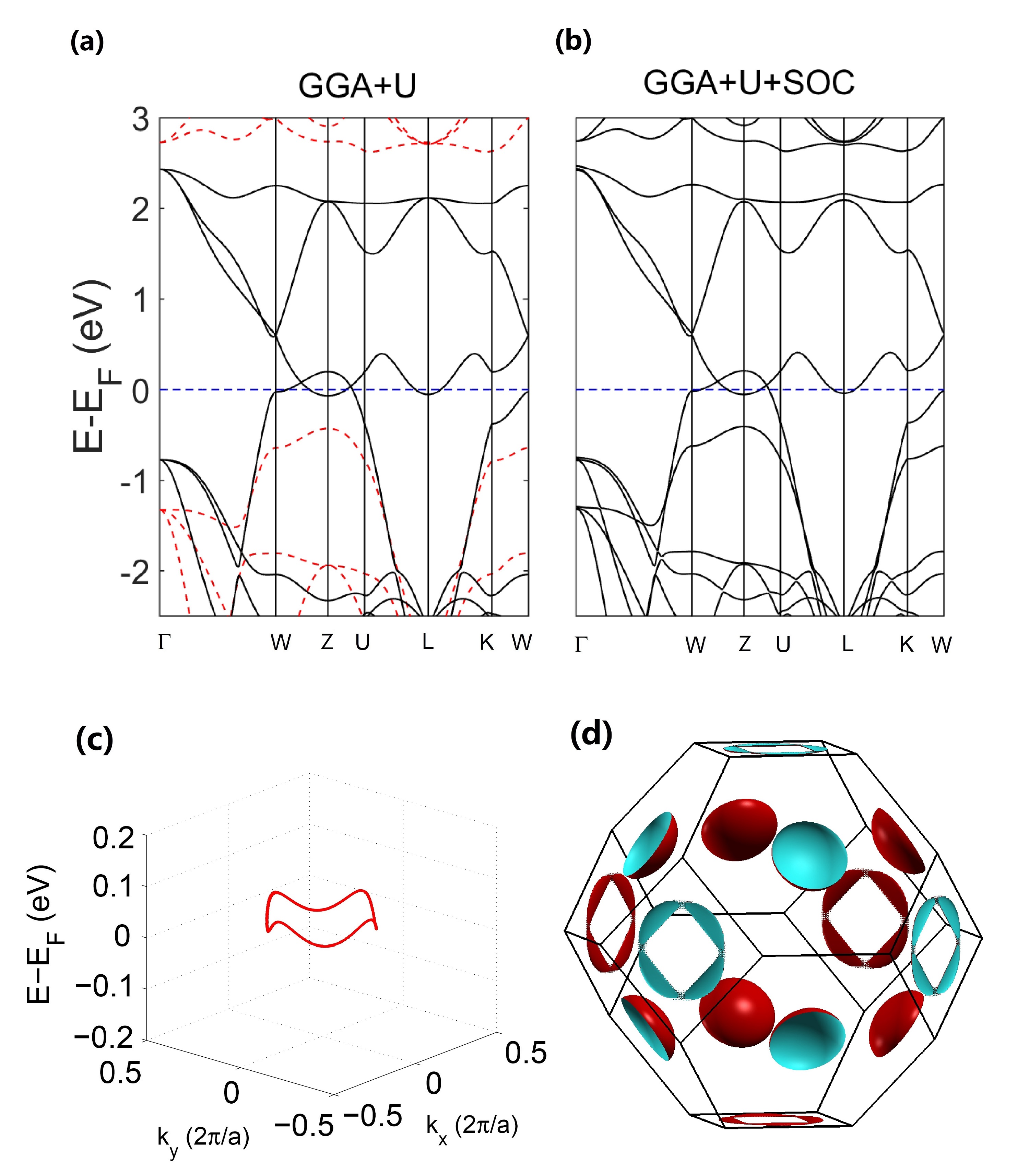}
	\caption{(a) The band structures of strained cubic CrO$_{2}$ at lattice constant $a = 1.02a_{0}$  without SOC and (b) with SOC. The majority and minority spin channels are indicated by the solid (black) and dashed (red) lines, respectively. (c) The energy dispersion of the DNLR in the BZ boundary of $k_{z}=2\pi/a$. (d) The 3D Fermi surface is calculated with a chemical potential of 30 meV in the presence of SOC. The DNR is located at the BZ boundary of $k_{z}=2\pi/a$. The approximate or accidental nodal rings occurs at boundaries of $k_{x}=2\pi/a$ and $k_{y}=2\pi/a$. \label{figure3}}
\end{figure}

Our results show that cubic CrO$_{2}$ presents FM phase with magnetic moment $~2\mu_{B}$ per unit cell and band gap $\sim$0.08 eV, which is insensitive to the SOC. The energy of FM state is about 56 meV lower than the antiferromagnetic state per Cr atom, indicating the high Curie temperature above room temperature of this compound. In the absence of SOC the FM order is isotropic, but the three principle axes are equally preferred after an arbitrarily small SOC is introduced, since electron orbits are anisotropic in a lattice. Without loss of generality, let us assume that the FM order is oriented along $\mathbf{z}$-direction, and look into its symmetries. The FM order reduces the point group from $O_h$ to $C_{4h}$, for which rotations are through $\mathbf{z}$-axis, and the $\mathbf{x}$-$\mathbf{y}$ plane is the mirror plane for the reflection. Combined with $T$, the ferromagnetic CrO$_{2}$ with SOC is symmetric under the magnetic group $D_{4h}(C_{4h})$. Note that $T$ must be applied to recover the direction of the FM order after it is reversed by a rotation of $\pi$ through an axis normal to it, for instance the two-fold rotation axes of $D_{4h}$, $(1,0,0)$, $(1,1,0)$, $(0,1,0)$ and $(1,-1,0)$ in our case. Particularly in the BZ boundary of $k_z=2\pi/a$ (which is of our main interest), each point has its own little group leaving the states above it invariant. The little group is $C_{s}$ for a generic point, $C_{2v}(C_{s})$ for a generic point in the two-fold axes of $D_{4h}$, and $D_{4h}(C_{4h})$ for point $Z$.
It is observed that all points in the boundary share the unitary mirror symmetry through the $k_x$-$k_y$ plane of $k_z=0$, which may protect a DNL through quantizing the Berry phase of valence bands along a gapped mirror-symmetric circle~\cite{Chiu-Reflection-Classification,Fang2015}.

We now focus on the band structure restricted in the BZ boundary of $k_z=2\pi/a$. In Fig. \ref{figure2}(a), we report that the cubic CrO$_{2}$ is transited to a semi-metallic phase under small stretches of the lattice constant. By increasing the tensile strain equally along three principle axes, two bands closest to the Fermi energy are inverted at the Z point, and the crossing points form a continuous nodal ring. The gap at point Z as a function of lattice constant is shown Fig. \ref{figure2}(b), and it is found that the nodal ring emerges when the negative hydrostatic strain exceeds the critical point $a_{c}\sim 1.006a_{0}$. In the main text we choose $a = 1.02a_{0}$, while the stability analysis at different strengths of the hydrostatic strain and the dependence on the correlated energy U are reported in the Supplementary Material (SM)~\cite{Supp}. Panel (a) of Fig. \ref{figure3} shows the spin-polarized band structures of the strained cubic CrO$_{2}$ with and without SOC. Here, we see a band gap $\sim$2.9 eV of the minority spin states, while the majority spin states show semi-metallic features. The crossing points along W-Z and Z-U directions are approximately at 45.9 meV and 23.2 meV above the Fermi level, respectively, forming hole pockets. Remarkably, as seen in Fig. \ref{figure3}(c), the spin-polarized nodal ring of the FM CrO$_2$ is nearly flat with only an energy variance about 23 meV, and resides very close to the Fermi level, which provides a better platform to explore novel DNL physics compared with already-known candidates~\cite{Yu2015prl,Bian2016nc,Bian2016prb}.

Due to the shared abelian symmetry group $C_s$ consisting of the reflection $M_{xy}$ and the identity on the plane, each band corresponds to either of the two irreducible spinful representations $\Gamma_3$ and $\Gamma_4$ of $C_s$, where $M_{xy}$ has eigenvalue $i$ for $\Gamma_3$ and $-i$ for $\Gamma_4$. The symmetry group for the high symmetry point Z is the abelian group $C_{4h}$, and each state at $Z$ corresponds to a spinful irreducible representation of $C_{4h}$. As shown in Fig. \ref{figure2}(a), when gapped the conduction and valence band closest to the Fermi energy correspond to $\Gamma_3$ ($\Gamma^+_{7}$ at $Z$) and $\Gamma_4$ ($\Gamma^-_{7}$ at $Z$), respectively, and therefore the two crossed bands have opposite eigenvalues of the mirror symmetry $M_{xy}$ (see the SM for character tables~\cite{Supp}). Thus the formed nodal ring is a DNR, and the Berry phase of the valence band along any mirror-symmetric circle surrounding it is quantized by the unitary mirror symmetry $M_{xy}$~\cite{Zak-phase}.

\begin{figure}
	\includegraphics[scale=0.35]{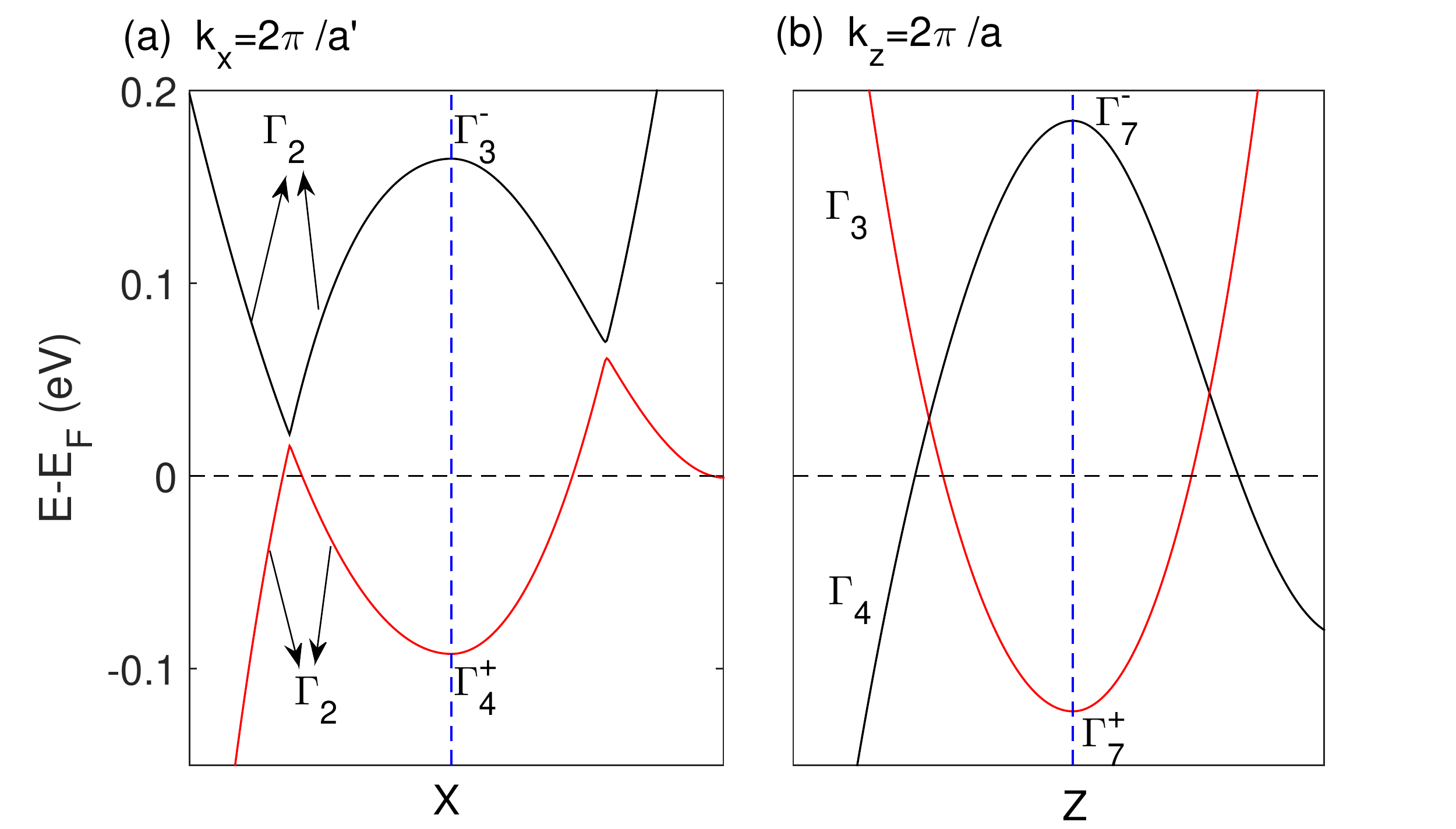}
	\caption{The stability of band crossings under anisotropic strains preserving $D_{4h}(C_{4h})$. The lattice constant $a'$ for the principle axes $\mathbf{x}$ and $\mathbf{y}$ is one percent longer than the lattice constant $a$ for the axis $\mathbf{z}$, namely $(a'-a)/a=0.01$.\label{Anisotropic-Strain}}
\end{figure}

Under the strain equal for three principle directions, it looks in Fig. \ref{figure3}(d) that nodal rings appear also in the BZ boundaries of $k_x=2\pi/a$ and $k_y=2\pi/a$, for which the degeneracy, however, are accidental or approximate beyond the numerical resolution. These nodal rings, if accidentally exist, can be eliminated by symmetry-preserving perturbations, since the mirror symmetries for these planes are anti-unitary, and therefore cannot quantize the Berry phase along a mirror-symmetric circle~\cite{Supp}. This is demonstrated by exerting one percent extra tensile strain equally along $\mathbf{x}$ and $\mathbf{y}$ directions, which preserves the $D_{4h}(C_{4h})$ symmetry of the FM phase. As shown in Fig. \ref{Anisotropic-Strain}, the energy gap is now clearly present, while the DNR on the BZ boundary of $k_z=2\pi/a$ still robustly exists.

\begin{figure}
	\centering
	\includegraphics[scale=0.09]{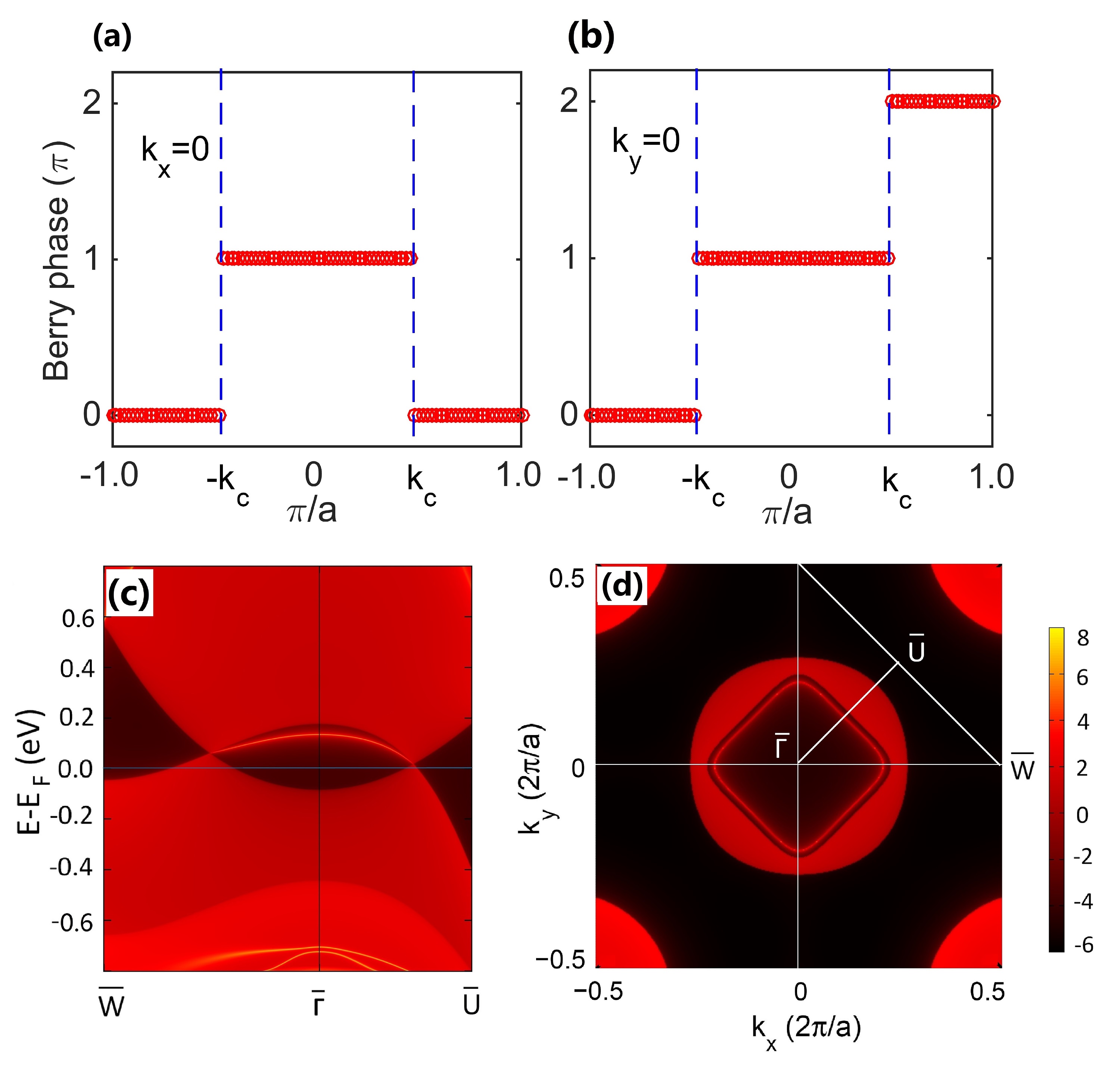}
	\caption{Quantized Berry phases along mirror-symmetric circles and the induced drumhead surface states. (a) and (b) illustrate the distribution of Berry phases for 1D subsystems along lines $k_x=0$ and $k_y=0$, respectively. (c) Local density of states projected into the (001) surface of strained cubic CrO$_{2}$ with SOC.
	(d) The projected Fermi surface is calculated with a chemical potential of 30 meV. Red solid regions represent the bulk states, and the thin red circle represents the intersection with the drumhead surface state. \label{Berry-surface}}
\end{figure}

We now proceed to discuss the drumhead surface states originated from the mirror-quantized Berry phase of the DNR. Since the eigenvalues of $M_{xy}$ for the crossed bands forming the ring are opposite to each other, the Berry phase of the valence band cumulated from a mirror-symmetric circle surrounding the DNR
is quantized to be $\nu\equiv\pi\mod 2\pi$, noting that a mirror-symmetric gauge transformation over the circle may change the Berry phase by an integer multiples of $2\pi$~\cite{Zak-phase,Supp}. The chosen circle can be continuously deformed in a mirror symmetric way into two 1D $k_z$-subsystems with one inside the DNR and the other outside.
Then we have the relation among Berry phases, $\nu\equiv N_R-N_L\mod 2\pi$, where $N_R$ and $N_L$ are quantized Berry phases of the valence band for 1D subsystems inside and outside the DNR, respectively. Thus if 1D subsystems inside the DNR has nontrivial Berry phase ($\pi\mod 2\pi$), those outside must have trivial Berry phase ($0\mod 2\pi$), and vice versa. Our numeric result illustrated in Figs. \ref{Berry-surface}(a) and (b) shows that 1D subsystems side the DNR have nontrivial Berry phase. From the bulk-boundary correspondence, there should be drumhead surface states covering the region inside the projection of the nodal ring on the surface BZ, which is confirmed by our numerics. To obtain the surface states, we construct a tight-binding Hamiltonian with the basis of maximally localized Wannier functions \cite{ Marzari2012, Mostofi2008}, and thereby employ the method of Green's function \cite{Sancho1985}. Panel (c) of Fig. \ref{Berry-surface} shows the calculated local density of states, and (d)
Fermi surfaces for a semi-infinite (001) surface of strained cubic CrO$_{2}$ in the presence of SOC. Drumhead surface states cover the internal region of the projection of the DNR on the surface BZ. These surface states are unoccupied, and are located in the vicinity of the Fermi level with a tiny band-width of about 30 meV.


\textit{Short Summary--}The cubic CrO$_2$ under the anisotropic tensile strains with the magnetic symmetry $D_{4h}(C_{4h})$ has been identified by first-principles calculations as a mirror-symmetry-protected ferromagnetic DNLSM at room temperature. Considering that CrO$_2$ is a very common material in experiment and the DNR close to the Fermi energy is ideally flat, our work paves the way for further explorations on the novel physic and applications of DNL and drumhead fermions, particularly those related to their magnetic properties.

\begin{acknowledgments}
	\emph{Acknowledgments--}
This work is supported by the National Natural Science
Foundation of China (NSFC, Grant Nos.11204185, 11304403,
11334003 and 11404159).
\end{acknowledgments}
~~~\\
\textbf{AUTHOR INFORMATION}\\
$^*$\textbf{Equal Contributions:}\\
R. Wang and J. Z. Zhao contributed equally to this work.\\
$^{\dag}$\textbf{Corresponding authors:}\\
yuxinfruit@gmail.com (Y.X.Z.), xuh@sustc.edu.cn (H.X.), tong.sy@sustc.edu.cn (S.Y.T)\\

\bibliographystyle{apsrev}
\bibliography{FMDNL-Ref}

\clearpage
\newpage

\appendix

\setcounter{figure}{0}
\makeatletter

\makeatother
\renewcommand{\thefigure}{S\arabic{figure}}
\renewcommand{\thetable}{S\Roman{table}}
\renewcommand{\theequation}{S\arabic{equation}}
\begin{center}
	\textbf{
		\large{Supplemental Material for}}
	\vspace{0.2cm}
	
	\textbf{
		\large{
			``Ferromagnetic Dirac-Nodal Semimetal Phase of Stretched Chromium Dioxide" }
	}
\end{center}

\vspace{-0.2cm}

In this supplemental Material, we present the stability analysis at different hydrostatic strain, the band structures as a function of correlated energy U, the character tables of irreducible spinful representations for $C_{4h}$ and $C_s$, and the quantization of the Berry phase by mirror symmetries of the magnetic group $D_{4h} (C_{4h})$.
\section{The phonon of cubic chromium dioxide at different values of hydrostatic strain}
\begin{figure}
	\centering
	\includegraphics[scale=0.06]{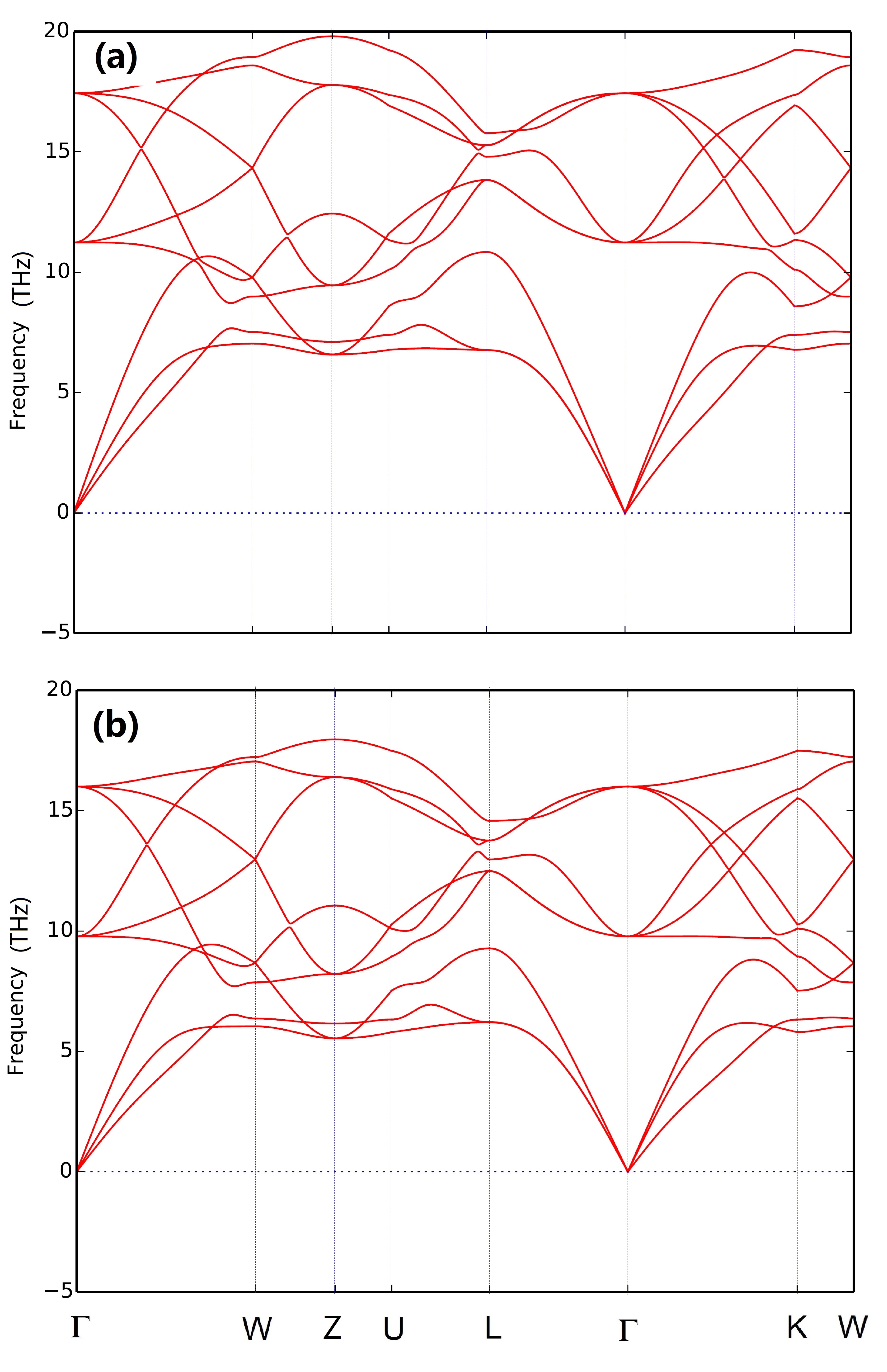}
	\caption{The phonon dispersions of cubic CrO$_{2}$  at $a=a_{0}$ and $a=1.02a_{0}$. \label{figS1}}
\end{figure}

The phonon spectrum is one useful way to investigate the stability and structural rigidity.  The method of force constants has been used to calculate
the phonon frequencies as implemented in PHONOPY package~\cite{Togo1,Togo2,Togo3}. We employ
$4 \times4 \times 4$ supercell with 64 Cr atoms and 128 O atoms to obtain the real-space force constants. Our results for the phonon dispersions at $a=a_{0}$ and $a=1.02a_{0}$ are shown in panels (a) and (b) of Fig.\ref{figS1}, respectively. We find that there is the absence of any imaginary frequencies over the entire BZ, demonstrating that the cubic CrO$_{2}$ and its strain structure are dynamical stability.

\section{The band structures as a function of the correlated energy}
\begin{figure}
	\centering
	\includegraphics[scale=0.09]{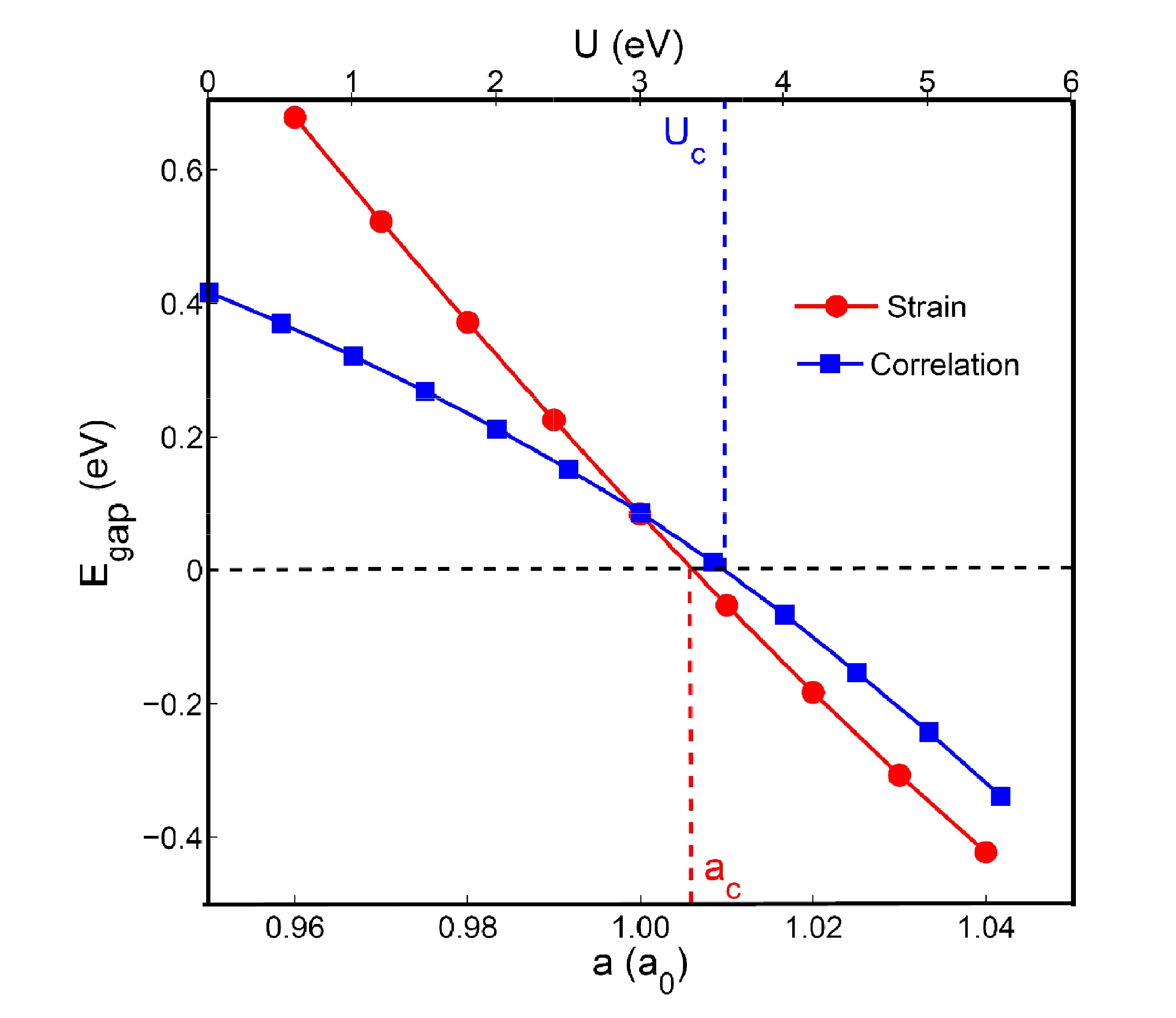}
	\caption{Energy gap of cubic CrO$_2$ at the Z point as a function of electronic correlation U (blue) under the ambient pressure. The negative gap implies the band inversion. For comparison, the energy gap is plotted as a function of lattice constant at U=3.0 eV.  \label{figS2}}
\end{figure}

It is interesting that the procedure of band inversion has also been found by tuning the on-site correlated energy U. With increasing the correlated energy, the evolution behaviors of electronic band structures is quite similar to those of the tensile hydrostatic strain.  As shown in Fig. \ref{figS2}, the band inversion occurs at $U_c = 3.55$ eV under the ambient pressure.

\section{The character tables for irreducible spinful representations for $C_{4h}$ and $C_s$}
The groups $C_{4h}$ and $C_s$ are represented projectively as subgroups of $O(3)$ due to the spin degrees of electrons, noting that rotating a spin by $2\pi$ through any axis produces an extra phase $-1$. Both $C_{4h}$ and $C_s$ are abelian, and therefore have only one-dimensional irreducible representations. Accordingly each row of the character tables is simply an irreducible spinful representation of $C_{4h}$ (or $C_s$), and all the rows in the table gives a complete list of irreducible spinful representations of $C_{4h}$ (or $C_s$).

\begin{table}
	\renewcommand\arraystretch{1.8}
	\begin{tabular}{p{0.8cm}|*{1}{p{0.75cm}} *{8}{p{0.75cm}} }
		\hline
		$C_{4h}$      & \centering $E$ &\centering  $C_4$          & \centering $C_2$ & \centering $C_4^3$        & \centering $i$ & \centering $S_4^3$       & \centering $s_h$ & $S_4$ \\
		\hline
		$\Gamma_5^{+}$ & \centering 1   &\centering $\omega$       &\centering $i$   &\centering $\omega ^{-1}$ &\centering 1  &\centering $\omega$       &\centering $i$   & $\centering \omega^{-1}$ \\
		$\Gamma_6^{+}$ & \centering 1   &\centering $\omega ^{-1}$ &\centering $-i$  &\centering $\omega$       &\centering 1  &\centering $\omega ^{-1}$ &\centering $-i$  & $\centering\omega$ \\
		$\Gamma_7^{+}$ & \centering 1   &\centering $-\omega$      &\centering $i$   &\centering $-\omega ^{-1}$&\centering 1  &\centering $-\omega$      &\centering $i$   & $\centering -\omega ^{-1}$ \\
		$\Gamma_8^{+}$ & \centering 1   &\centering $-\omega ^{-1}$&\centering $-i$  &\centering $-\omega$      &\centering 1  &\centering $-\omega ^{-1}$& \centering$-i$  & $\centering -\omega$ \\
		$\Gamma_5^{-}$ & \centering 1   &\centering $\omega$       &\centering $i$   &\centering $\omega ^{-1}$ &\centering -1 &\centering $-\omega$      &\centering $-i$  & $\centering -\omega ^{-1}$ \\
		$\Gamma_6^{-}$ & \centering 1   &\centering $\omega ^{-1}$ &\centering $-i$  &\centering $\omega$       &\centering -1 &\centering $-\omega ^{-1}$&\centering $i$   & $\centering -\omega$ \\
		$\Gamma_7^{-}$ & \centering 1   &\centering $-\omega$      &\centering $i$   &\centering$-\omega ^{-1}$ &\centering -1 &\centering $\omega$       &\centering $-i$  & $\centering \omega ^{-1}$ \\
		$\Gamma_8^{-}$ & \centering 1   &\centering $-\omega ^{-1}$&\centering $-i$  &\centering $-\omega$      &\centering -1 &\centering $\omega ^{-1}$ &\centering $i$   & $\centering \omega$ \\
		\hline
	\end{tabular}
	\caption{The character table of irreducible spinful representations for the point group $C_{4h}$. For the row of group elements, $E$ is the identity, and $i$ is the inversion. $C_4$, $C_2$, $C_4^3$ are rotations of $\pi/4$, $pi/2$ and $3\pi/4$, respectively, through the principle axis. $S_4=iC_4$, $S_4^3=iC_4^3$,$s_h=iC_2$ that is the reflection through the plane perpendicular to the principle axis. In the bulk of the table, $i=e^{i\pi/2}$ and $\omega=e^{i\pi/4}$.}
\end{table}

\begin{table}
	\renewcommand\arraystretch{1.8}
	\begin{tabular}{p{0.8cm}|*{1}{p{2.0cm}} *{2}{p{0.75cm}} }
		\hline
		$C_{s}$ & \centering$E$ & $s$ \\
		\hline
		$\Gamma_{3}$ &\centering 1 & i \\
		$\Gamma_{4}$ &\centering 1 & -i \\
		\hline
	\end{tabular}
	\caption{The character table of irreducible spinful representations for the point group $C_{s}$. Here $E$ is the identity, $s$ is the reflection through a plane, and $i=e^{i\pi/2}$.}
\end{table}

\section{The quantization of the Berry phase by mirror symmetries of the magnetic group $D_{4h} (C_{4h})$}

\textit{Unitary mirror Symmetry--}Without loss of generality, we consider a one-dimensional lattice model, for which the mirror symmetry is just an inversion. Considering a single band theory, the Bloch wave function is accordingly transformed as
\begin{equation}
M|k\rangle=ie^{i\theta(-k)}|-k\rangle, \label{Mirror-Operation}
\end{equation}
where the phase is periodic in the Brillouin zone zone, $e^{i\theta(-\pi)}=e^{i\theta(\pi)}$. By including only one band, we have implicitly assumed that the spins of all electrons are aligned perpendicular to the mirror plane, and therefore the mirror operation does not flip spins. Applying the mirror reflection twice simply multiply the wave function of the spin by $-1$, namely
\begin{equation}
M^2=-1,
\end{equation}
which when applied to Eq.(\ref{Mirror-Operation}) requires
\begin{equation}
e^{i\theta(-k)}=e^{-i\theta(k)}. \label{Unitary-M-Condition}
\end{equation}
In particular at mirror-invariant points $k=0$ or $\pi$, the phase is $e^{i\theta(0~ \mathrm{or}~\pi)}=\pm 1$.

Now we show that the mirror symmetry can quantize the Berry phase accumulated by moving the Block function over the whole Brillouin zone.
\begin{equation}
\begin{split}
N&=\int_{-\pi}^\pi dk~ \langle k|i\partial_k |k\rangle\\
&=\int_{-\pi}^\pi dk~ \langle k|M^\dagger i\partial_k M|k\rangle\\
&=\int_{-\pi}^\pi dk~ \langle -k|e^{-i\theta(-k)}i\partial_k e^{i\theta(-k)}|-k\rangle\\
&=-\int_{-\pi}^\pi dk~ \langle k|e^{-i\theta(k)}i\partial_k e^{i\theta(k)}|k\rangle\\
&=-N-\int_{-\pi}^\pi dk~ e^{-i\theta(k)}i\partial_k e^{i\theta(k)}\\
&=-N+2\pi n_W[e^{i\theta(k)}], \label{DerivationBerry}
\end{split}
\end{equation}
where $n_W$ is the winding number of $e^{i\theta(k)}$ as a map from $S^1$ to $U(1)$. So the Berry phase on the mirror-symmetric 1D circle can only take value in integer multiples of $\pi$, namely $N=\pi n_W[e^{i\theta[k]}]$. It is noteworthy that the condition of Eq. (\ref{Unitary-M-Condition}) from the unitary mirror symmetry allows a nonvanishing winding number.

In addition, a gauge transformation in the whole BZ, $|k\rangle\rightarrow e^{i\phi(k)}|k\rangle$, changes the Berry phase by an integer multiple of $2\pi$, $2\pi n_W[e^{i\phi(k)}]$. Thus we define the mirror-symmetry protected topological invariant as the gauge-invariant quantity,
\begin{equation}
N\equiv \int_{-\pi}^\pi dk~ \langle k|i\partial_k |k\rangle\mod 2\pi.
\end{equation}
Historically this was first pointed out by Zak \cite{Zak}.

\begin{figure}
	\includegraphics[scale=0.07]{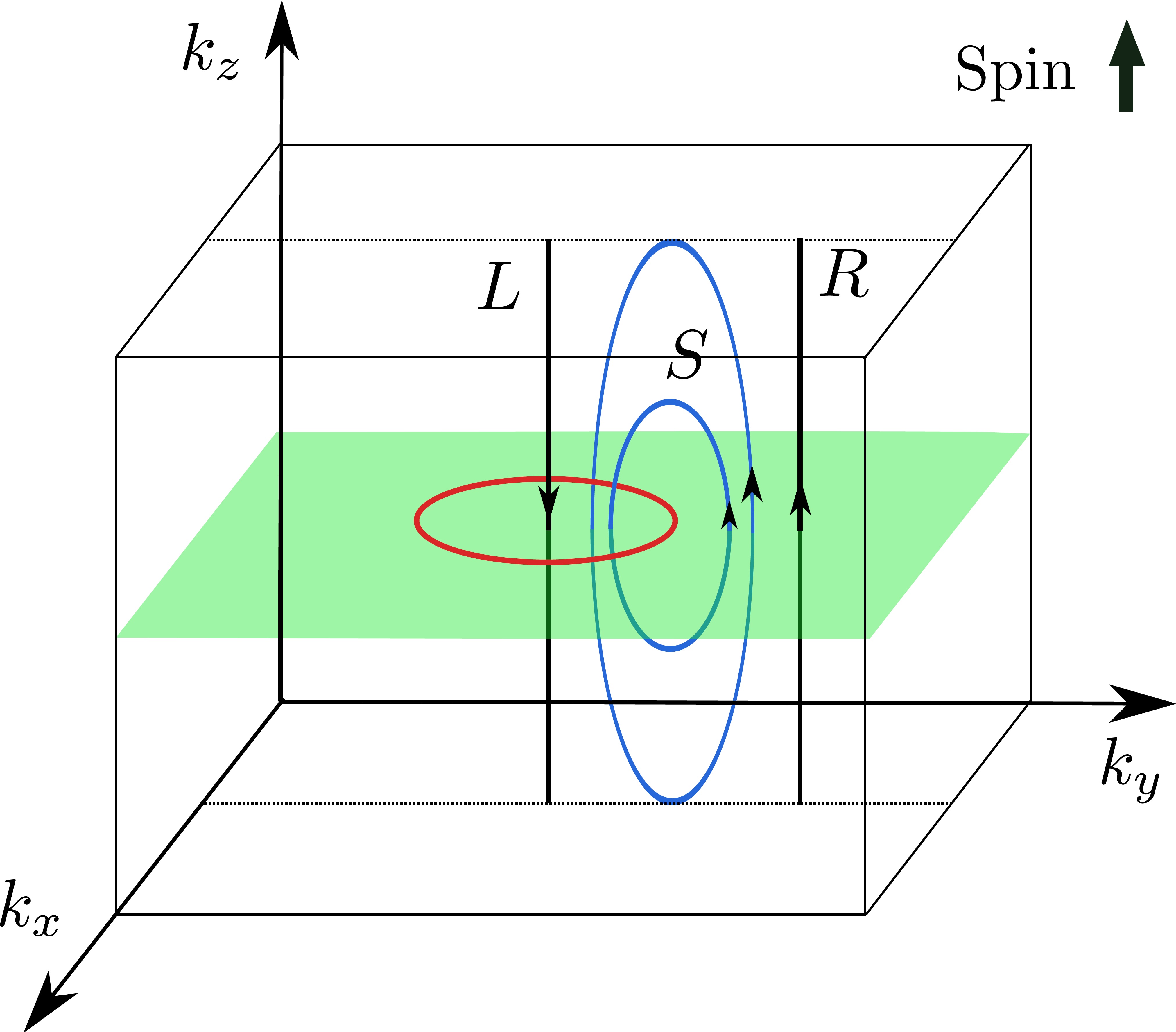}
	\caption{ A mirror-symmetric small circle is continously deformed in a mirror-symmetric way into two 1D subsystems. The FM order is aligned along $\mathbf{z}$-direction, and accordingly the mirror plane is the $k_x$-$k_y$ plane perpendicular to the FM order. The DNR is denoted by the red circle in the mirror-invariant plane. \label{figS3}}
\end{figure}

Here, we assume that the FM order is oriented along $\mathbf{z}$-direction. Since the eigenvalues of $M_{xy}$ for the crossed bands forming the ring are opposite to each other, the Berry phase of the valence band cumulated from any mirror-symmetric circle, for instance $S$ in Fig. \ref{figS3}, is quantized to be $\nu\equiv\pi\mod 2\pi$, noting that a mirror-symmetric gauge transformation along the circle may change the Berry phase by any integer multiples of $2\pi$. The small circle $S$ locally surrounding the nodal ring may be continuously deformed mirror symmetrically to be two 1D subsystems parametrized by $k_z$ with one inside the ring and the other outside as illustrated in Fig. \ref{figS3}.

\textit{Antiunitary mirror symmetry--}
There is also anti-unitary mirror symmetries in the magnetic group $D_{4h}(C_{4h})$, which are the combinations of T with a two-fold rotations along principle axes normal to the magnetic order, namely $TC_{2\mathbf{x}}$ and $TC_{2\mathbf{y}}$ with the magnetic order along $\mathbf{z}$-direction. Such an anti-unitary mirror symmetry $\tilde{M}$ satisfies
\begin{equation}
\tilde{M}^2=1,\quad \tilde{M}i\tilde{M}=-i.
\end{equation}
A band as
\begin{equation}
\tilde{M}|k\rangle=e^{i\tilde{\theta}(-k)}|-k\rangle.
\end{equation}
The applying $\tilde{M}$ twice, it is found that
\begin{equation}
e^{i\tilde{\theta}(k)}=e^{i\tilde{\theta}(-k)},
\end{equation}
which implies that the winding number of $e^{i\tilde{\theta}(k)}$ as a function from $S^1$ to $U(1)$ must be zero, namely $n_W[e^{i\tilde{\theta}(k)}]=0$. Similar to Eq. (\ref{DerivationBerry}), we can find $N=\pi n_W[e^{i\tilde{\theta}(k)}]=0$, which means the Berry phase always vanishing in order to preserve the anti-unitary mirror symmetry. Thus Berry phase is not a topological invariant in this case.

Form another viewpoint, a gapped mirror-symmetric circle with anti-unitary mirror symmetry can actually be regarded as a one-dimensional topological insulator with time-reversal symmetry in the symmetry class AI, which is well known to be topologically trivial.

\section{Mirror symmetric deformation of a circle into two 1D subsystems}
As illustrated in Fig. \ref{figS3}, a small mirror-symmetric circle $S$ enclosing the DNR is continuously deformed in a mirror-symmetric way, first to be a large circle with its lower and upper points being related by a reciprocal translation, and then into two 1D subsystems parametrized by $k_z$. Since the mirror symmetry is preserved in the entire process and the therefore the Berry phase is well quantized, we have the identity of Berry phases for valence bands,
\begin{equation}
\nu=N_R-N_L\mod 2\pi,
\end{equation}
where $\nu$ is the Berry phase for the small circle $S$, and the $N_{R}$ ($N_{L}$) correspond to the 1D $k_z$-system inside (outside) the DNR.

\end{document}